\documentclass[10pt,final,Twcolumn]{IEEEtran}
\usepackage[cmex10]{amsmath}
\usepackage{adjustbox}
\usepackage{tabularx}
\usepackage{subfig}
\usepackage{tabularx}
\usepackage{graphicx}
\usepackage{float}
\usepackage{amssymb}
\usepackage{epstopdf}

\DeclareGraphicsExtensions{.eps,.pdf,.png}

\usepackage[switch]{lineno}

\begin{document}
%

%
\title{Data Analysis in Multimedia Quality Assessment: Revisiting the Statistical Tests}

\author{Manish~Narwaria,
 Luk\'{a}\v{s}~Krasula, and
      Patrick~Le Callet
\thanks{Manish Narwaria is with Dhirubhai Ambani Institute of Information and Communication Technology (DA-IICT), Gandhinagar, Gujarat, 382007, India. Luk\'{a}\v{s}~Krasula and Patrick Le Callet are with LS2N/IPI group, University of Nantes, 44306, France
e-mail: (manish\_narwaria@daiict.ac.in, lukas.krasula@univ-nantes.fr, patrick.lecallet@univ-nantes.fr).}
\thanks{}}


\maketitle



%

\begin{abstract}
Assessment of multimedia quality relies heavily on subjective assessment, and is typically done by human subjects in the form of preferences or continuous ratings. Such data is crucial for analysis of different multimedia processing algorithms as well as validation of objective (computational) methods for the said purpose. To that end, statistical testing provides a theoretical framework towards drawing meaningful inferences, and making well grounded conclusions and recommendations. While parametric tests (such as $t$ test, ANOVA, and error estimates like confidence intervals) are popular and widely used in the community, there appears to be a certain degree of confusion in the application of such tests. Specifically, the assumption of normality and homogeneity of variance is often not well understood. Therefore, the main goal of this paper is to revisit them from a theoretical perspective and in the process provide useful insights into their practical implications. Experimental results on both simulated and real data are presented to support the arguments made. A software implementing the said recommendations is also made publicly available, in order to achieve the goal of reproducible research.
\end{abstract}

\section{Introduction}
The growth of low-cost devices has virtually made multimedia signals an integral part of our daily lives. Todays end users are constantly interacting with multimedia, and are more demanding in terms of their multimedia experience, and perceptual quality is one of the intrinsic factors affecting such interaction. As a result, assessment of perceptual quality is an important aspect in todays multimedia communication systems \cite{ITU-SG}. The most reliable way of quality estimation typically involves the use of a human subject panel who provides ratings/preferences for the targeted multimedia content \cite{ITU-SG}, \cite{ITURBS1534}. This is referred to as subjective assessment. In contrast, objective estimation of quality relies on the use of computational (mathematical) models \cite{ITUtutorial} that are expected to mimic subjective perception.

\begin{figure*}
\centering
{ \includegraphics[width=.7\textwidth]{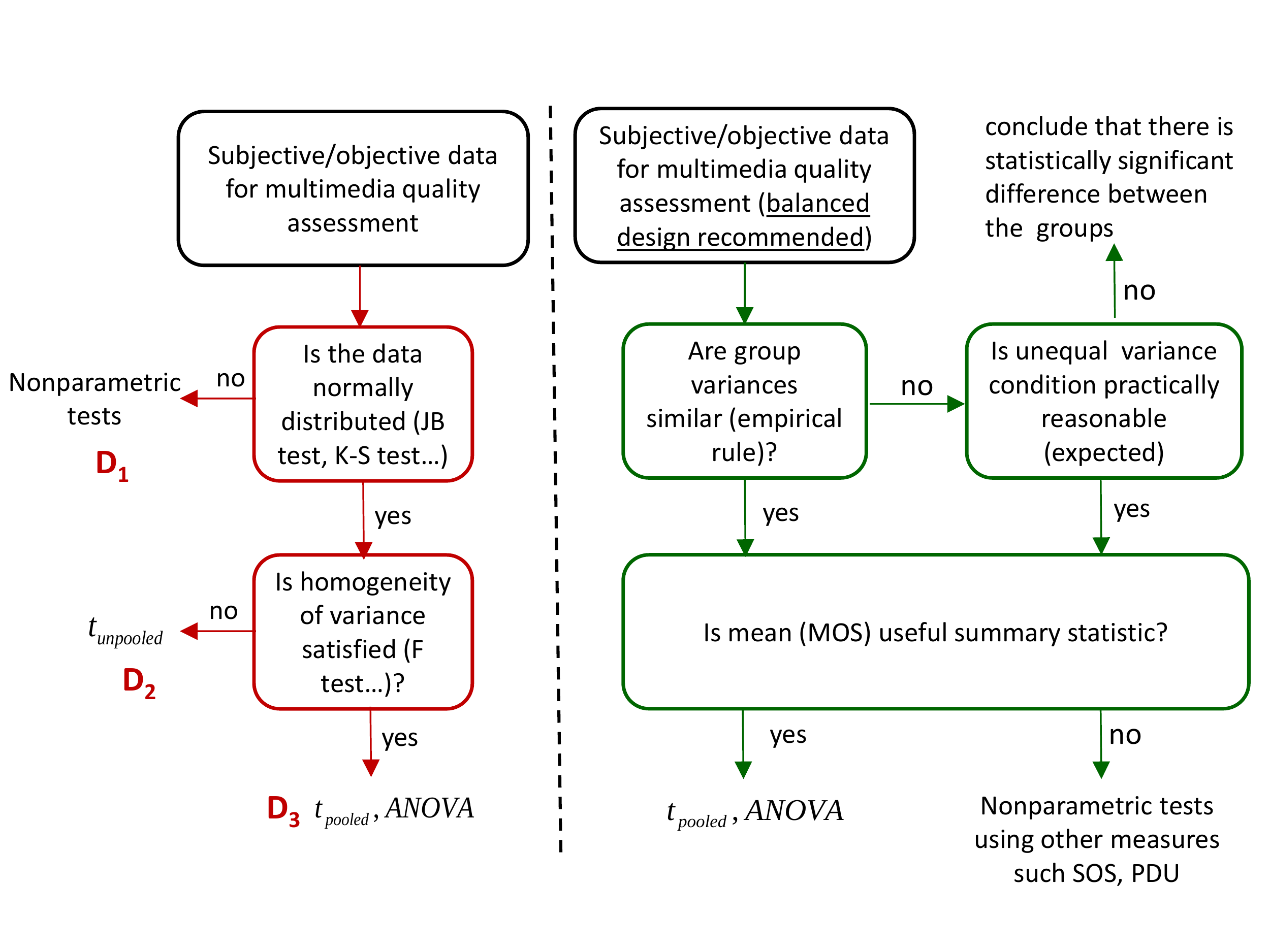}}
\caption{Typical procedure of applying parametric tests (the left flow chart) and the recommended approach (right flow diagram). The drawbacks associated with making decisions D\textsubscript{1}, D\textsubscript{2} or D\textsubscript{3} are discussed in sections \ref{assumption of normality} and  \ref{homogeneity of variance}. Figure best viewed in color.}
\label{dectree}
\end{figure*}

Parametric statistical tests find extensive application in multimedia quality estimation mainly for two purposes. First, they are used to compare and analyze subjective data collected from human participants. For instance, a $t$-test can be used to compare Mean Opinion Score (MOS) from two different conditions in a variety of applications (eg. analyzing codec performance \cite{7254155}, investigating the effect of upscalers on video quality \cite{Pitrey}, studying optimization criteria in HDR tone mapping \cite{doi:10.1117/1.OE.52.10.102008} and so on). Analysis of Variance (ANOVA) is also a commonly used technique for analyzing the effect of two or more factors/treatment levels and their interactions. These include identifying audiovisual interactions \cite{belmudez2016audiovisual}, examining the impact of reflections in HDR video tone mapping \cite{Melo2015}, investigating the effect of resolution, bit rate and color space on under water videos \cite{underwaterVQA}, studying the possible impact of compression level and type of content on perceptual quality towards finding optimal presentation duration in subjective quality assessment \cite{optimal_duration} etc. Second, these tests are used to validate objective (computational) methods against subjective data. This can in turn be used to statistically compare several objective methods in terms of their prediction accuracies as compared to the subjective data. Such validation studies are obviously central to benchmarking objective methods before they can be deployed in practice.

The need for statistical testing arises due to the fact that subjective studies use a finite sample of human subjects. Therefore, these tests can help in generalizing and making inferences for the population. For that purpose, parametric tests such as $t$-test, $F$-test, ANOVA, and error estimation (eg. using confidence intervals) are widely used in the community. While the application of parametric tests is generally straightforward (aided by the availability of numerous software packages), the interpretation of the results requires some care. In particular, statistical tests in many cases are simply treated as {\it {black boxes}}, and are applied without considering the practical implications of the assumptions in these tests. 

As the name implies, such tests are based on {\it{apriori}} knowledge of parameterizable probability distribution functions (eg. $t$ distribution, $F$ distribution which are respectively characterized by one and two degrees of freedom.). While it is true that parametric tests are distribution dependent (as opposed to non-parametric tests which are some times referred to as being {\it{distribution-free}}), there appears to be some confusion regarding the assumptions made in these tests. In particular, the assumption of normality and homogeneity of variance  in many cases appears to be not well understood for both subjective and objective data analysis. In practice, these assumptions are sometimes considered as bottle necks in applying parametric tests. As a result, nonparametric tests are recommended if the data violates one or both the assumptions. A typical approach to applying parametric statistical tests is depicted in the left flow diagram in Figure \ref{dectree}, and consists of arriving at one of the three decisions D\textsubscript{1}, D\textsubscript{2} or D\textsubscript{3}:
\begin{itemize}
\item D\textsubscript{1}: normality checks (eg. JB test, K-S test) are applied to examine if the given subjective/objective data is normal. If such normality checks determine the data to be {\it{nonnormal}} then nonparametric tests are carried out.

\item D\textsubscript{2}: If the normality test determines the data to be {\it{normal}}, then homogeneity of variance is tested by applying a test of variance (eg. Levene's test, $F$ test etc). If the groups/samples do not satisfy the said assumption then modified tests (eg. unpooled $t$ test) are applied which do not use pooled variance in computing the test statistic.

\item D\textsubscript{3}:  If the data satisfies both assumptions of normality and homogeneity of variance then the usual $t$ test or ANOVA (which employ pooled variance) are applied.

\end{itemize}

In this paper, we seek to draw attention to few drawbacks associated with such decisions. Specifically, we revisit theoretical formulations and the resultant practical implications to highlight shortcomings and recommend alternative approach (right flow diagram in Figure \ref{dectree}) in the light of the said assumptions. We emphasize that these assumptions should not be viewed as constraints or bottle necks in the application of parametric tests. Instead these should be carefully considered and understood in the context of their practical implications. Subsequently, we provide a set of recommendations to ameliorate some of the drawbacks that may stem from either wrong interpretation or application of the said assumptions in parametric testing. A software implementing the said recommendations is also made publicly available\footnote{https://sites.google.com/site/narwariam/home/research}, in order to achieve the goal of reproducible research.

The remainder of the paper is organized as follows. In section \ref{assumption of normality} we analyze the distributional assumptions in parametric test. Section \ref{homogeneity of variance} provides an analysis of the assumption of homogeneity of variance. Section \ref{practical considerations} points out the practical implications in the context of multimedia quality assessment. In section \ref{experiments} we present the experimental results and analysis while Section \ref{recommendations} lists a set of recommendations towards proper use of parametric testing in the context of the said assumptions. We provide concluding thoughts in section \ref{concluding remarks}.

\section{Revisiting distributional assumptions in parametric tests} \label{assumption of normality}

Parametric tests require certain assumptions including the assumption of normality, homogeneity of variance and data independence. As highlighted in left flow diagram in Figure \ref{dectree}, normality checks have usually been applied on subjective or objective data \cite{ITURBS1534}, \cite{ITUtutorial}, \cite{7254155}, \cite{P1401}. Such use of normality checks indicates that the assumption of normality is, in many cases, misunderstood to be applicable on the data for which statistical tests are to be carried out. This is, however, incorrect in the light of the fact that all parametric tests essentially work by locating the observed test statistic on a known probability distribution function. Then, depending on the desired significance level and the location of test statistic, one typically accepts or rejects the null hypothesis. For example, in $t$-test, the $t$-statistic is first computed from the observed sample. This $t$-statistic is then compared with values from a $t$-distribution (corresponding to the particular degrees of freedom).  In other words, the computed test statistic ($t$-statistic, $F$-statistic etc.) is assumed to follow the corresponding distribution ($t$-distribution in $t$-test, $F$-distribution in $F$-test and ANOVA etc.). 

Thus, the more appropriate question to be asked in parametric testing is whether the test statistic follows the assumed distribution (rather than the data being normally distributed). The answer to such question requires that the subjective (or objective) test be repeated for a large number of times, each time using a different sample (both in terms of human subjects and content). Then, in each instance, the test statistic can be computed to obtain its sampling distribution.  This process is, however, neither practical for obvious reasons nor desirable. Instead, one can rely on the fundamental central limit theorem (CLT). Informally, the CLT states that the sampling distribution of the arithmetic mean (and sum) will approach a normal distribution as the sample size increases, regardless of the underlying population distribution \cite{PFLUG1983323}. It is due to this result that the test statistic in parametric tests are guaranteed to follow the assumed distribution, provided that the sample size is large enough (approaching infinity in theory). 

We begin by considering two populations ${\it\bf  {p_1}}$ and ${\it\bf  {p_2}}$ with means $\mu_1$ and $\mu_2$ and variances $\sigma_1^2$ and $\sigma_2^2$, respectively. In the context of multimedia quality assessment, these populations will typically represent the collection of subjective (or objective) opinion scores for two conditions (eg. subjective or objective quality scores for two profiles of a video codec, indivudual quality scores for audiovisual content corresponding to two parameter settings, quality scores for content rendered by two depth image based rendering methods, individual quality scores for two tonemapped HDR videos and so on) for which we need to compare mean quality scores i.e. $\mu_1$ and $\mu_2$. Assume that ${\it\bf  {p_1}}$ and ${\it\bf  {p_2}}$ are sampled i.e. subjective or objective assessment is actually performed on a set of content using a sample of human subjects or using objective methods. Let the corresponding samples be denoted by ${\it\bf  {x_1}} = [x_{11},...,x_{1n_1}]$ and ${\it\bf  {x_2}} = [x_{21},...,x_{2n_2}]$ where $n_1$ and $n_2$ are the sample sizes, and the sample observations are assumed to be independent and identically distributed (iid) random variables. Note that there are no assumptions regarding the distribution of either the populations (${\it\bf  {p_1}}$ and ${\it\bf  {p_2}}$) or corresponding samples (${\it\bf  {x_1}}$ and ${\it\bf  {x_2}})$.

\subsection{Sampling distribution of test statistic in $t$-test}

Let $\overline {x}_1$, $\overline {x}_2$ and $s_1^2$, $s_2^2$ denote the sample means and variances, respectively. Then the goal of the analysis is to infer if $\mu_1 = \mu_2$ (the {\it{null}} hypothesis) or not. To that end, one can employ the $t$-test. To define the $t$-statistic, we use the result from the CLT i.e.
\begin{equation}  
	\overline {x}_1 \sim N\left(\mu_1, \frac{\sigma_1}{\sqrt{n_1}}\right) \text{and} \; \overline {x}_2 \sim N\left(\mu_2, \frac{\sigma_2}{\sqrt{n_2}}\right)
\end{equation}
Then, the difference between the samples means will also be normally distributed i.e.
\begin{equation}  \label{eq:2}
	\overline {x}_1 - \overline {x}_2 \sim N\left(\mu_1-\mu_2, \sqrt {\frac{\sigma_1^2}{{n_1}}+\frac{\sigma_2^2}{{n_2}}}\right) 
\end{equation}
By standardization, we have
\begin{equation}  \label{eq:3}
	\frac{\overline {x}_1 - \overline {x}_2 - (\mu_1-\mu_2)}{\sqrt {\frac{\sigma_1^2}{{n_1}}+\frac{\sigma_2^2}{{n_2}}}} \sim N\left(0,1\right) 
\end{equation}  
Note that in eq. (\ref{eq:3}) only the numerator is a random variable while the denominator is constant. However, in practice, the population variance is generally not known. We therefore need to use sample variance as an {\it{unbiased estimator}} of the population variance. To proceed further, we consider two cases for defining the {\it{null}} hypothesis.
\subsubsection{Case 1: Samples drawn from same population}
We can define the {\it{null}} hypothesis as $H_0:$ the two samples are taken from the same population. This implies that not only are we assuming the population means to be equal but other population parameters including variances are equal. Thus, we have $\mu_1 = \mu_2$ and $\sigma_1^2 = \sigma_2^2 =\sigma^2 $ (say). In order to obtain a more accurate estimate of the (common) population variance, we can employ the pooled variance $s_p^2$ which is defined as

\begin{equation} \label{eq:6}
	s_p^2 =  { {\frac{s_1^2(n_1-1)+s_2^2(n_2-1)}{{(n_1-1)+(n_2-1)}}}} 
\end{equation}

Thus, under $H_0$, the denominator in eq. (\ref{eq:3}) can be modified accordingly and the $t$-statistic defined as

\begin{equation} \label{eq:7}
	t_{pooled} = \frac{\overline {x}_1 - \overline {x}_2}{s_p\sqrt {\frac{1}{{n_1}}+\frac{1}{{n_2}}}}, df_{pooled} = n_1+n_2-2
\end{equation}

With the said modification, the reader will now note that the denominator in eq. (\ref{eq:7}) is also a random variable, unlike eq. (\ref{eq:3}) where it was a constant. Thus, $t_{pooled}$ is a ratio of two random variables. The numerator is the difference of two independent normally distributed random variables ($\overline {x}_1$ and $\overline {x}_2$), and will therefore be normally distributed \cite{roussas2003introduction}. Further, the squared denominator will be equal to ${\frac{s_p^2}{{n_1}}+\frac{s_p^2}{{n_2}}}$ which denotes the variance of the said normal distribution in the numerator. Hence, the denominator in eq. (\ref{eq:7}) will be chi-squared distributed \cite{roussas2003introduction}. Accordingly, the test statistic $t_{pooled}$ is characterized by the ratio of normally and square root of chi-squared distributed variables. It will therefore be approximately\footnote{In theory, the sample size should tend to infinity for the sample means to be normally distributed according to CLT. However, in practice, smaller samples sizes allow us to approximate the assumption of normality, regardless of population or sample distribution.} distributed according to the $t$-distribution \cite{roussas2003introduction} with $df_{pooled} = n_1+n_2-2$ degrees of freedom, and this will be irrespective of the distribution of either the populations (${\it\bf {p_1}}$ and ${\it\bf  {p_2}}$) or corresponding samples (${\it\bf  {x_1}}$ and ${\it\bf  {x_2}})$.
\subsubsection{Case 2: Samples drawn from two different populations with same population mean}

In the second case, we assume that the two samples have been drawn from two different populations with same population mean i.e. $\mu_1 = \mu_2$ (but $\sigma_1^2 \neq \sigma_2^2$). Hence, other population parameters such as variance or any other statistic need not be equal.  Then, we can use sample variances as an estimate of the two population variances, and under the assumption of the {\it{null}} hypothesis, eq. (\ref{eq:3}) can be modified to obtain the following test statistic

\begin{equation} \label{t_unpooled}
	t_{unpooled} = \frac{\overline {x}_1 - \overline {x}_2}{\sqrt {\frac{s_1^2}{{n_1}}+\frac{s_2^2}{{n_2}}}}, \: df_{unpooled} = \frac{\left(\frac{\sigma_1^2}{n_1}+\frac{\sigma_2^2}{n_2} \right)^2}{{\frac{\sigma_1^4}{n_1^2(n_1-1)}+\frac{\sigma_2^4}{n_2^2(n_2-1)}}} 
%
\end{equation}

In practice, we use $s_1^2$ and $s_2^2$ to compute $df_{unpooled}$ in eq. (\ref{t_unpooled}) because $\sigma_1^2$ and $\sigma_2^2$ are not known. We will discuss the two cases in section \ref{to pool or not to pool}.

\subsection{The case of ANOVA and $F$-test}

The sampling distribution of the test statistic ($F$) in $F$-test (ANOVA also relies on $F$-test) is assumed to follow the $F$-distribution \cite{roussas2003introduction}. It can be shown that this assumption is valid irrespective of the data distribution with the same caveat concerning the CLT mentioned in the previous sub-section. 

Before doing that, we assume that there are $k$ groups each with $n_i$ observations (let the total number of observations be denoted by $M = \displaystyle\sum_{i=1}^{k}n_i$), and define the following: mean $\overline {x}_i$ of $i^{th}$ group, grand mean $\overline {X}$ and variance $s_i^2$ of the $i^{th}$ group as
\begin{equation} \label{eq:8}
\overline {x}_i = \frac{\displaystyle\sum_{j=1}^{n_i}x_{ij}}{n_i}, s_i^2 = \frac{\displaystyle\sum_{j=1}^{n_i}\left(x_{ij}-\overline {x}_i\right)^2}{n_i}, \overline {X} = \frac{\displaystyle\sum_{i=1}^{k}\displaystyle\sum_{j=1}^{n_i}x_{ij} }{\displaystyle\sum_{i=1}^{k}n_i} 
%
\end{equation}


The $F$-statistic in ANOVA is defined as the ratio of inter-group (i.e. between groups) and intra-group (i.e. within each group) variations. We denote these quantities by $SS_B$ and $SS_W$, respectively, with the corresponding degrees of freedom being $df_B$ and $df_W$. Then, the $F$-statistic is computed as
\begin{equation}  \label{eq:9}
	F = \frac{SS_B/df_B}{SS_W/df_W} = \frac{\displaystyle\sum_{i=1}^{k} n_i\left(\overline {x}_i-\overline {X}\right)^2/\left(k-1\right)}{\displaystyle\sum_{i=1}^{k}\displaystyle\sum_{j=1}^{n_i} \left(x_{ij}-\overline {x}_i\right)^2/(M-k)}  
\end{equation}
By noting that the denominator in eq. (\ref{eq:9}) is essentially a weighted sum of individual group variances, we can view the $F$-statistic as  
\begin{equation} \label{eq:10}
	F =  \frac{\displaystyle\sum_{i=1}^{k} n_i\left(\overline {x}_i-\overline {X}\right)^2/\left(k-1\right)}{\frac{n_1s_1^2+n_2s_2^2+...+n_ks_k^2}{(n_1-1)+(n_2-1)...+(n_k-1)}}  
\end{equation}
One can see that the numerator in eq. (\ref{eq:10}) is squared difference of two normally distributed variables ($\overline {x}_i$ and $\overline {X}$), and will be thus chi-squared distributed. The denominator can be seen to be very similar to the pooled variance used in eq. (\ref{eq:6}), and will be chi-squared distributed following similar arguments. It follows that $F$ is a ratio of two chi-squared distributed random variables which in turn implies that it will be approximately distributed according to the $F$-distribution (with $k-1$ and $M-k$ degrees of freedom). Once again, this is independent of the distribution of the population or the groups, and only relies on the approximations related to sample size as required in the CLT. 

\subsection{Data normality checks: are they required?} \label{multiple testing}
As discussed in previous sub-sections, the CLT being a theoretical result only provides asymptotic approximation in that as sample size tends to infinity the sampling distribution of mean tends to be normally distributed, and this holds irrespective of the sample or population distribution \cite{PFLUG1983323}. Note that the CLT does not specify any sample size above which the said sampling distribution will be normal. In practice, smaller sample sizes are generally sufficient to allow reasonable approximations. For instance, in the context of subjective quality assessment, Ref. \cite{15_obs} recommends a minimum of 15 subjects while the authors in \cite{24_obs} suggested using at least 24 subjects for audiovisual quality measurement. Because the sampling distribution of mean is directly or indirectly used in computing the test statistics such as $t$, $F$ etc., there are no requirements of normality (or any other distribution) on the data to be analyzed. It is, therefore, not surprising that previous works \cite{ITURBS1534}, \cite{schmider2010}, \cite{tskewed} have noted that parametric tests such as ANOVA are {\it{robust to non-normal data distributions}}, and the focus on distributional assumptions in these tests is not required \cite{1b98f554a52645909840a63ab0bad7a3}.

The second theoretical argument against the application of normality checks before conducting parametric tests is the inflation of Type I error probability. A commonly adopted strategy is to first check whether the given sample/data is normally distributed or not. To that end, normality tests such as the Kolmogorov-Smirnov (K-S) test, Jarque-Bera test, Shapiro-Wilk test etc. are popular. If the tests determine the given data is normally distributed then a parametric test is used. Otherwise, a non-parametric test is performed. As a result of this two-step process, there will be an increase in type I error probability. Assume that $H_0^*:\text {given data is normally distributed}$ (the {\it{null}} hypothesis in a normality test) and $H_0$ be the {\it{null}} hypothesis of the test that will follow. Then, the probability of rejecting $H_0$ can be written as the sum of mutually exclusive events i.e.

\begin{align} \label{multiple_tests}
P({\text {reject}} \:H_0) = P({\text {reject}} \:H_0 {\text {\:and not reject}}\: H_0^*)   \nonumber \\ + P({\text {reject}} \:H_0 {\text {\:and reject}}\:H_0^*) 
\end{align}

In the above equation the first expression on right hand side corresponds to the case of using a parametric test while the second expression corresponds to the use of a suitable non-parametric test. Because the critical regions corresponding to the parametric and non-parametric tests will be in general different, the resultant critical region which is a union of the critical regions of the individual tests is increased. Consequently, the probability\footnote{This probability value is not related to the $p$ value of the significance test. Instead, it refers to the probability (over repeated trials) of making a type I error i.e. rejecting $H_0$ when it is true.} to reject $H_0$ (when it is true) is increased thereby increasing the probability of a type I error. 

The third argument against the use of normality tests is the theoretical contradiction concerning the sample size. It is known that most normality tests, by definition, tend to reject the {\it{null}} hypothesis $H_0^{*}$ ({given data is normally distributed}) as the sample size increases. For instance, in the JB test for normality, the test statistic value is directly proportional to the sample size. In other words, larger the sample size, it is more likely to be determined as non-normal. However, according to CLT, the approximation of normality of the sampling distribution of mean improves as the sample size increases. This leads to a contradiction between the requirement of data normality and the asymptotic behavior in the CLT. 

While other methods such visual (eg. histogram visualization, normal probability plots) or those based on empirical rules (eg. if sample kurtosis is between 2-4, then the sample is deemed to be normally distributed) can overcome the limitations associated with the more formal normality tests, these are not required because it is the normality of sampling distribution of mean that is needed rather than the data being normal.

\section{To pool or not to pool?} \label{to pool or not to pool} \label{homogeneity of variance}
      
In this section, we analyze the assumption of homogeneity of variance and point out the theoretical aspects that need to be considered in the context of this assumption. The relevant practical considerations will be discussed in the next section.

\subsection{Should homogeneity of variance be checked?}

As discussed in the previous section, the {\it{null}} hypothesis can be defined in two cases. For Case 1, we require the assumption of homogeneity of variance (i.e. $\sigma_1^2 = \sigma_2^2$) and is applicable in the context of ANOVA (for more than two groups) and $t_{pooled}$ (for two groups). Note that both the tests use an estimate of the pooled variance in order to compute the corresponding test statistic. On the other hand, Case 2 does not require homogeneity of variance and is applicable in defining the test statistic $t_{unpooled}$. Therefore, $t_{unpooled}$ is widely used in statistical data analysis and has been included in many statistical packages such as SPSS. However, it can be noted that in general $df_{unpooled} < df_{pooled}$ (except when $\sigma_1^2 = \sigma_2^2$ and $n_1 = n_2$, in which case both are equal), and hence the use of $t_{unpooled}$ will increase the probability of Type II error (i.e. the test will be more conservative). In light of this, a popular and seemingly logical strategy is to first conduct a preliminary test of variance based on which a decision to either use $t_{pooled}$ (or ANOVA) or $t_{unpooled}$ (if the test of variance leads to the conclusion that $\sigma_1^2 \neq \sigma_2^2$). 

Notice that this strategy, however, involves cascaded use of the given data in rejecting or accepting two hypotheses (one from test of variance and the other from the $t$-test). In other words, two significance tests are performed on the same data. As a consequence, the Type I error probability will be increased \cite{roussas2003introduction}. Suppose $H_0^{**}:\sigma_1^2 = \sigma_2^2$ (the {\it{null}} hypothesis in a preliminary variance test for equality of population variances) and $H_0:\mu_1 = \mu_2$ be (the {\it{null}} hypothesis for the $t$-test that will follow). Then, the probability of rejecting $H_0$ in this case can be written as (similar to eq. \ref{multiple_tests}) the sum of probability of rejecting $H_0$ when  $H_0^{**}$ is not rejected and the probability of rejecting $H_0$ when  $H_0^{**}$ is also rejected. Following the same arguments as in section \ref{multiple testing}, the resultant critical region which is a union of the critical regions of the individual $t$-tests is increased thereby inflating the probability of Type I error. 

Further, note from eq. (\ref{t_unpooled}) that the degrees of freedom for $t_{unpooled}$ depends on population variances $\sigma_1^2$ and $\sigma_2^2$, and will therefore be a random variable in case these are estimated from sample variances (which is practically the more likely case). As a result, its analysis, both theoretical and experimental is more complicated due to the fact that its distribution is not independent of sample variances \cite{welch}. Thus, the interest in $t_{unpooled}$ is more from a theoretical perspective in that it allows for a {\it{correction}} in degrees of freedom which in turn renders it valid in cases when population variances are not equal. In practice, however, it is more relevant to consider the implications of comparing means of two populations whose spread (variances) are different. Hence, applying statistical tests for checking homogeneity of variance prior to using $t$ test, ANOVA etc. is not recommended due to theoretical (due to increased probability of type I error) reasons, and is of less interest in practice. 

\subsection{The case of balanced design} \label{balanced design}
It can be shown that the test statistic $t_{pooled}$ is valid even if $\sigma_1^2 \neq \sigma_2^2$ provided that the sample sizes are equal (balanced design). To prove this, we compare the distributions of $t_{unpooled}$ and $t_{pooled}$ by writing them in terms of the theoretical $t$ distribution \cite{welch} in the following form:  
\begin{equation}
t_{pooled} = c_{pooled} \cdot t_{df_{pooled}},  t_{unpooled} = c_{unpooled} \cdot t_{df_{unpooled}}
\end{equation}
where $t_{df_{pooled}}$ and $ t_{df_{unpooled}}$ are the $t$ distributions with respective degrees of freedom. Thus, for $t_{pooled}$ and $t_{unpooled}$ to follow the respective theoretical $t$ distributions the corresponding multiplicative factors $c_{pooled}$ and $c_{unpooled}$ should be equal to 1. It can, however, be shown \cite{welch} that while $c_{unpooled} $ is always equal to 1,  the value of $c_{pooled} $ depends on sample size and population variances i.e.

\begin{equation}
c_{pooled} = \sqrt{\frac{\left(n_1+n_2-2\right)\left(\frac{\sigma_1^2}{n_1}+\frac{\sigma_2^2}{n_2} \right)}{\left(\frac{1}{n_1}+\frac{1}{n_1} \right)\{\left( n_1-1\right)\sigma_1^2+\left( n_2-1\right)\sigma_2^2   \}}}
\end{equation}  
From the above equation, it is easy to see that $c_{pooled} =1$ if the population variances are equal ($\sigma_1^2 = \sigma_2^2$). However, $c_{pooled}$ is also equal to 1 if sample sizes are equal ($n_1 = n_2$). In other words, $t_{pooled}$ will follow the expected theoretical distribution if balanced design is used, despite the violation of the assumption of homogeneity of variance. Because several practical applications tend to target a balanced design i.e. equal sample sizes, the use of $t_{pooled}$ is valid in such cases even if sample variances differ by a large amount. Particularly, in case of multimedia quality assessment, the use of balanced design is common. For instance, typical subjective quality assessment tests use the same number of human subjects to evaluate the quality of different conditions (although the subject panel may or may not comprise of the same subjects in evaluating the quality of each condition).

\section{Practical considerations in the domain of multimedia quality assessment} \label{practical considerations}

\begin{figure*}
\centering
\subfloat[]{ \includegraphics[width=.4\textwidth]{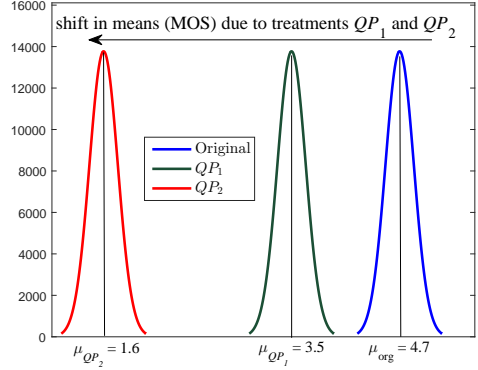}}
\caption{Illustration of treatment effects $QP_1$ and $QP_2$. The shift in location does not alter the variance of the groups. The values of $\mu_1$, $\mu_2$ and $\mu_3$ are assumed for illustration only. Figure best viewed in color.}
\label{hov}
\end{figure*}

In this section, we discuss the assumption of homogeneity of variance from the practical view point, and take an illustrative example from the domain of video quality assessment. Let us consider that an original (i.e. undistorted) video sequence is viewed and rated for its visual quality by all the concerned observers on a scale of 1 (worst) to 5 (excellent). Hence, this set of individual ratings forms the population of interest $P_{org}$ for this condition (i.e. undistorted video). We can express each element of $P_{org}$ as $P_{org}^{(i)}= {\mu_{org}+\epsilon_i}$ where $\mu_{org}$ is the mean of $P_{org}$ and $\epsilon_i$ denotes the random error (with zero mean and finite variance) that will be introduced in each individual rating. This error term can be used to take into account the fact that some observers may be more critical (so their corresponding ratings will be less than $\mu_{org}$) while others may be less critical (i.e. their ratings are expected to be higher than $\mu_{org}$) of the video quality. Suppose the said video is now compressed using two quantization parameter ($QP$) values $QP_1$ and $QP_2$ and $QP_2>QP_1$ ($QP$ is employed in video compression as a measure to quantify quantization levels, higher $QP$ implies higher quantization and in general lower video quality). 

\subsection{The case of systematic treatment effect}

In the considered example, quantization can be considred as a treatment that is applied to the original video. Assuming all other conditions to be identical (i.e. same display, ambient light, viewing distance etc.), the treatments $QP_1$ and $QP_2$ will decrease the video quality and essentially cause a shift in means (MOS). In other words, the intervention in original video will result in shifted (in location) version of the population $P_{org}$, as shown in Figure \ref{hov}. Let $\mu_{QP_1}$ and $\mu_{QP_2}$ denote the means of the populations $P_{QP_1}$ and $P_{QP_2}$, respectively. Then, if these treatments have a systematic effect on video quality, we can express the elements of the corresponding populations as $P_{QP_1}^{(i)}= {\mu_{org}+E_{QP_1}+\epsilon_i}$ and $P_{QP_2}^{(i)}= {\mu_{org}+E_{QP_2}+\epsilon_i}$. Here $E_{QP_1}$ and $E_{QP_2}$ are the effects of the treatments $QP_1$ and $QP_2$, respectively. Hence, the quality scores for the new conditions are shifted from $\mu_{org}$ by an amount triggered by the visible impact of the treatments on the video quality, and can be quantified by $E_{QP_1}$ and $E_{QP_2}$. In the example shown in Figure \ref{hov}, $E_{QP_1} = -1.2$ and $E_{QP_2} = -3.1$ (negative values are indicative of decrease in video quality). Notice that the resulting populations $P_{QP_1}$ and $P_{QP_2}$ will have the same variance as $P_{org}$ because the treatments ($QP_1$ and $QP_2$) will cause systematic changes in individual ratings (i.e. observers who were more critical in case of original video will remain so for the new conditions also). In the alternate case, if the treatments do not cause any changes in the opinion scores i.e the effect is not visible to the observers (i.e. $E_{QP_1} = 0$ and $E_{QP_2} = 0$), then the three populations will be the same and one can conclude that the treatments do not lead to statistically significant differences in means (MOS). 

\subsection{The case of heterogeneous variances} \label{heterogeneous variances}

In the third case, if the treatments $QP_1$ and $QP_2$ do not introduce systematic effect on video quality, then the individual opinion scores may randomly increase (video quality improves visibly according to some observers), decrease (video quality degrades visibly according to some observers) or remain the same (video quality levels remains same as without any treatment). In such case, we can say that the treatments caused the ratings to become heterogeneous because apart from the inherent random error ($\epsilon_i$), the varying values of $E_{QP_1}$ and $E_{QP_2}$ will introduce additional and possibly different variations in $P_{QP_1}$ and $P_{QP_2}$. Consequently, the variances of the three populations $P_{org}$, $P_{QP_1}$ and $P_{QP_2}$ will be different. Hence, testing if $\mu_{org} = \mu_{QP_1} = \mu_{QP_2}$ may not be useful since the populations will be different in any case. Practically, such cases are of less interest because one generally knows the effect of a given treatment {\it{apriori}} (in the given example of video compression, it is known $QP_1$ and $QP_2$ will lower video quality levels as compared to the original video) and statistical tests help to establish if the observed differences due to the treatment are merely due to chance (i.e. due to sampling error) or not. 

If the population variances are unequal, it may point out to 2 possibilities: (1) additional factors may have crept in, (2) the observers have not been consistent in their ratings. The first possibility is generally minimized by careful experimental design including training sessions at the beginning of the test to ensure that the participants have understood the task well. The effect of second possibility is mitigated by rejecting outliers i.e. inconsistent observers that can cause variance to change are removed from further studies or analysis. Such outlier rejection is well accepted and recommended in multimedia quality analysis, and well documented outlier rejection strategies exist \cite{15_obs}, \cite{VQEG}. Therefore, outlier rejection provides indirect support for the assumption of homogeneity of variance, even though the explicit goal is to remove data points which might be {\it{dissimilar}} rather than making the variances of groups similar. In other words, experimental design in subjective tests for quality will help to ensure that the variances of the groups to be analyzed are similar. In general, the issue of heterogeneous group variances can be avoided \cite{Sawilowsky2013FermatSE} if proper experimental guidelines have been followed. In other words, Case 2 (i.e. samples/groups drawn from different populations with same population mean) may be practically less useful although it is perfectly valid for theoretical analysis. In summary, careful experimental design is more crucial for reliable statistical analysis and comparisons rather than focusing on homogeneity of variance and/or distributional assumptions (data normality).

It may also be noted that while the use of $t_{pooled}$, ANOVA requires that population variances are equal, it does not imply that sample/group variances be exactly equal. Rather the said variances should be similar. This can be quantified by computing the ratio of maximum to minimum group variance. Empirically, if the said ratio is greater than or less than $1/4 \: (= 0.25$), then the population variances can be deemed to be unequal. In such case, it may not be meaningful to conduct $t$-test or ANOVA because the samples are likely to be drawn from two different populations. 

\subsection{Comparing groups with different variances}

Homogeneity of variance condition should be viewed in the light of practical considerations and not as a constraint. Therefore, it can be assessed via the empirical rule in order to obtain information about the presence of groups/samples that may have very different variances as compared to the remaining ones, and might suggest the possibility that the samples are taken from different populations (in which case comparing the means via $t_{unpooled}$ or other test which does not use pooled variance may be less meaningful). Once again, practical context should be used to ascertain if unequal variance condition is reasonable in view of the goals of analysis. For instance, it is possible that only a fraction of groups may violate this condition in which case the possible reasons can be examined. In other cases, such groups could possibly be removed from analysis. As discussed in section \ref{balanced design}, in theory $t_{pooled}$, ANOVA are in any case not affected by unequal variance if balanced design (equal sample size) is employed. Therefore experimental design should target balanced design as far as possible (in multimedia quality estimation, balanced design are common). Nevertheless, practically it may be more insightful to analyze the possible reasons and consequences of unequal variance rather than merely applying the statistical tests.

As discussed, Case 2 is valid from a theoretical perspective but is of less interest in practice. In other words, the implications of comparing $k$ samples whose corresponding populations have different variances but with equal means i.e. $\mu_1 = \mu_2 = ...=\mu_k$, should also be noted. In this context, it is useful to point out that MOS is sometimes not the most accurate measure of multimedia quality, and other measures may be required to supplement it. For instance, the authors in \cite{SOS} proposed the use of SOS (standard deviation of opinion scores) while Ref. \cite{PDU} suggested using PDU (percentage dissatisfied users) in addition to MOS. Note that measures such as SOS, PDU can be different even if corresponding population MOS are equal. Such cases will arise if groups (samples) from different populations (with same population means) are compared, and may not lead to meaningful analysis of perceptual quality and/or user satisfaction levels. 

\section{Experimental results and discussion} \label{experiments}

In the first set of experiments, we investigate the effect of type of distribution that the sample follows. We considered four different types of distributions (from which random numbers were generated to simulate sample observations), and these are summarized in Table \ref{table1}. Note that the parameters for these distributions were chosen in order to result in diverse shapes (in terms of symmetry, number of peaks etc.). The kurtosis values reported in Table \ref{table1} reflect this. 

\begin{figure*} [h]
\centering
\subfloat[Beta]{\label{as_psnr} \includegraphics[width=.4\textwidth]{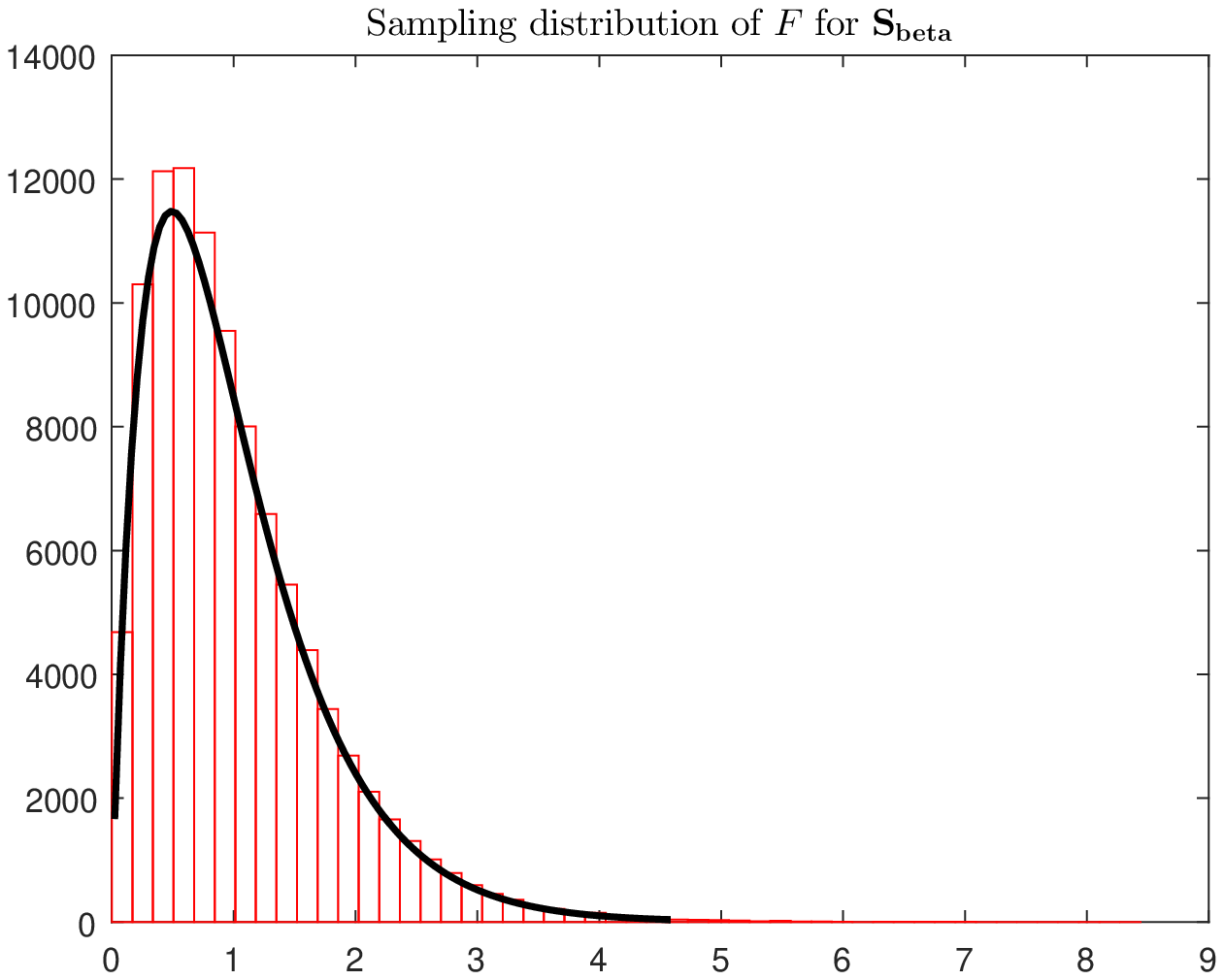}}
\hfil
\subfloat[Exponential]{\label{as_psnr} \includegraphics[width=.4\textwidth]{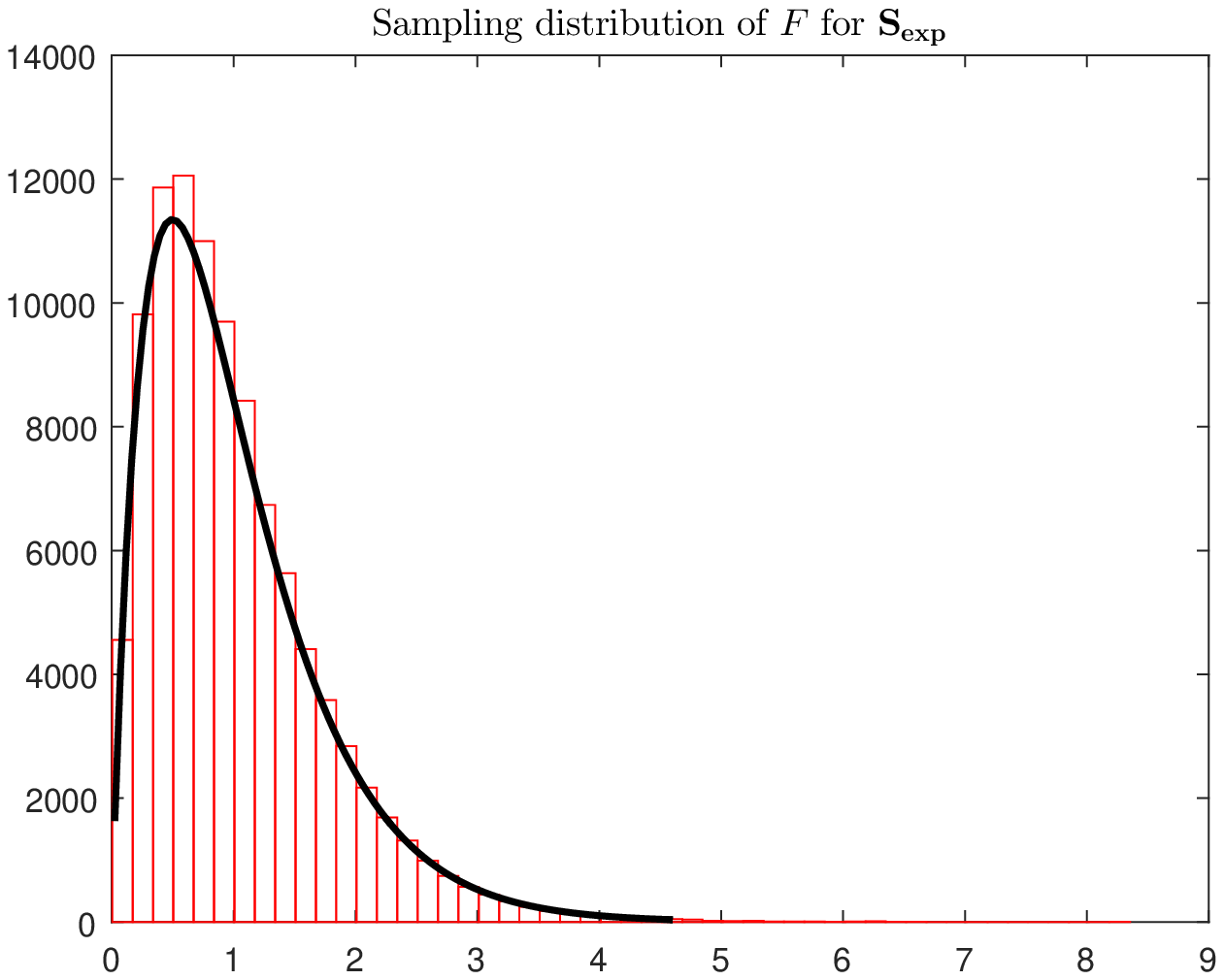}}
\hfil
\subfloat[Normal]{\label{as_psnr} \includegraphics[width=.4\textwidth]{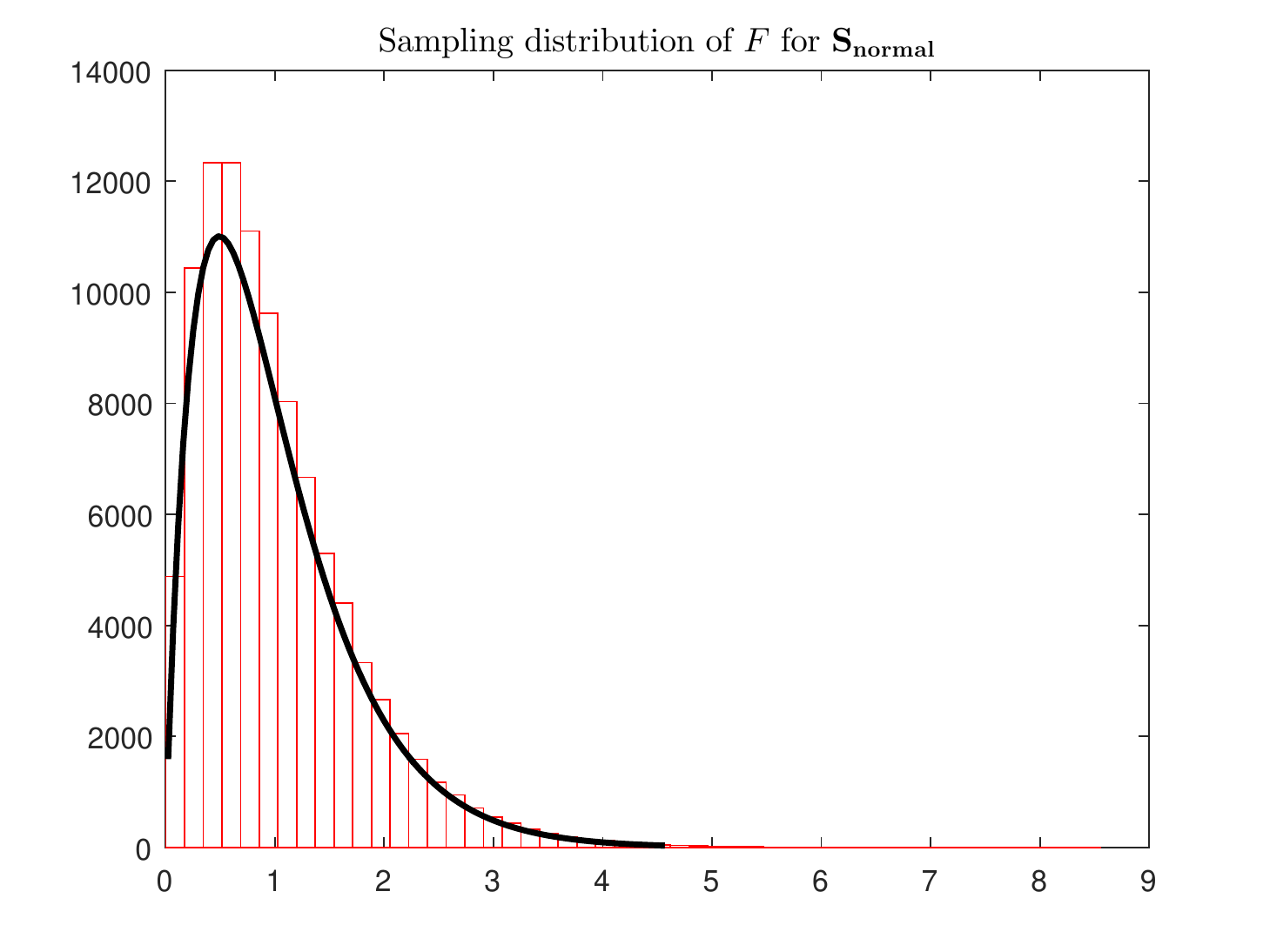}}
\hfil
\subfloat[Uniform]{\label{as_psnr} \includegraphics[width=.4\textwidth]{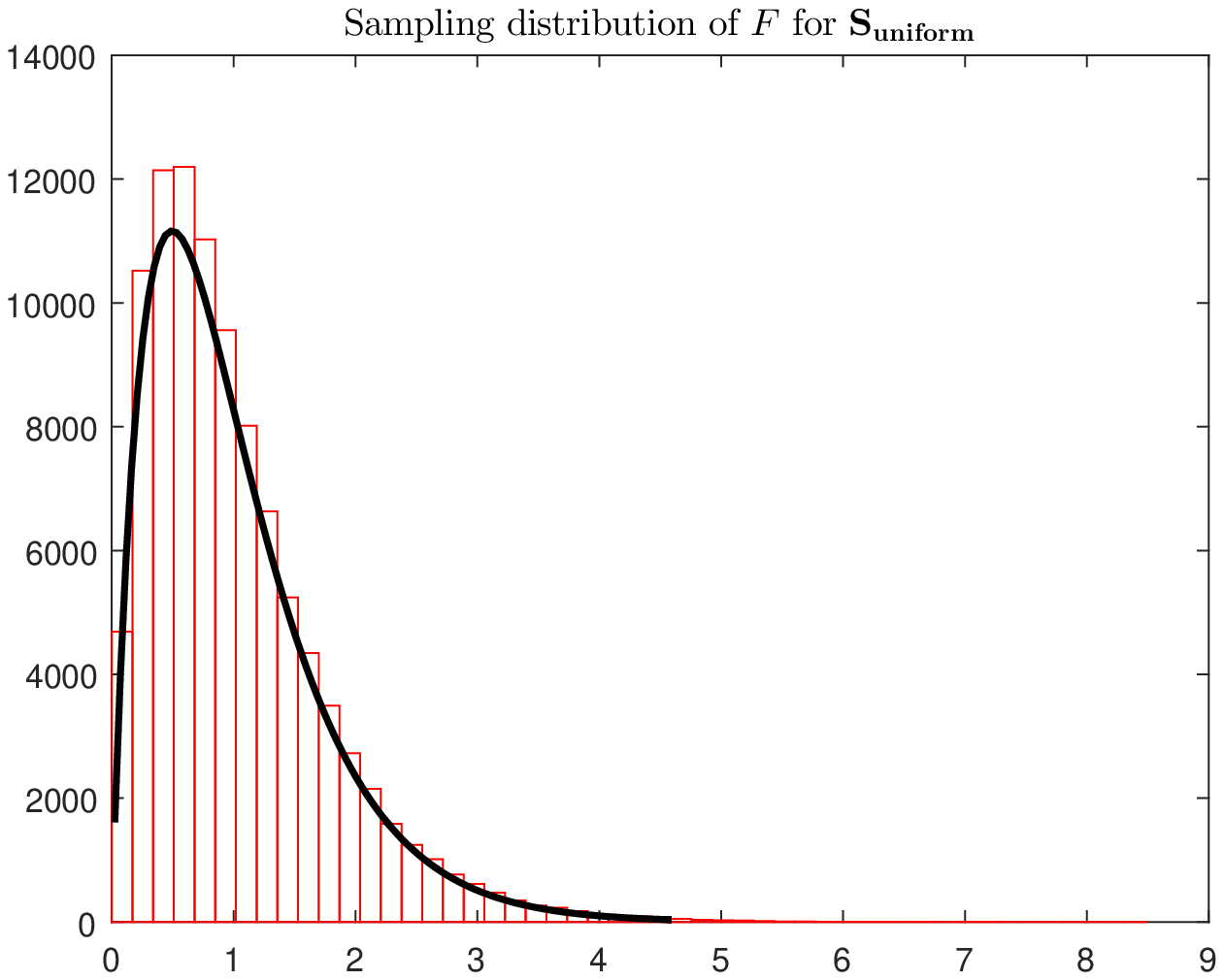}}
\caption{Sampling distribution of $F$ values when the samples follow the indicated distributions. In each plot, the continuous curve indicates the theoretical $F$-distribution with 4 and 120 degrees of freedom. Figure best viewed in color.}
\label{sampling distribution of F}
\end{figure*}

\begin{table} 
\caption{Description of distribution types and their characteristics.}
\label{table1}

\begin{tabular}{|c||c| |c| |c| }
\hline
Type  & Parameters & Shape & Kurtosis\\
\hline
Beta & \shortstack{$a=0.5$,\\$b=0.5$} & \shortstack{symmetric, \\bimodal (two peaks)}& 1.5  \\
\hline
Exponential & $\lambda=0.5$ & \shortstack{decaying curve, \\non-symmetric} & 9 \\
\hline
Normal & \shortstack{$\mu=0$,\\$\sigma=1$} & \shortstack{bell-shaped, symmetric, \\ unimodal (one peak)}   & 3 \\
\hline
Uniform & \shortstack{$a=0$,\\$b=1$} & \shortstack{flat (no peaks), \\symmetric} & 1.8 \\
\hline
\end{tabular}
\end{table}

As an example, we use ANOVA, and study the sampling distribution of $F$ when the samples follow the distributions mentioned in Table \ref{table1}. We consider 5 groups ($k=5$), equal number of observations in each group ($n_i = n = 25$), and ensured that the groups have similar variances. Thus, we represent the sample for exponential distribution as ${\bf{S}}_{exp} = [{\bf {d}}_{1exp} \ {\bf {d}}_{2exp} \ {\bf {d}}_{3exp} \ {\bf {d}}_{4exp}\  {\bf {d}}_{5exp}]$. Here ${\bf {d}}_{1exp}$ to ${\bf {d}}_{5exp}$ are 25-dimensional column vectors representing the groups. Similarly, we can define the samples for other distributions i.e. ${\bf{S}}_{beta}$, ${\bf{S}}_{normal}$ and ${\bf{S}}_{uniform}$.

Since our goal was to study the sampling distribution of $F$ in ANOVA, ${\bf{S}}_{exp}$, ${\bf{S}}_{beta}$, ${\bf{S}}_{normal}$ and ${\bf{S}}_{uniform}$ were generated randomly in each iteration, making sure the that observations followed the respective distributions. The sampling distributions of $F$ for each case are shown in Figure \ref{sampling distribution of F}. The number of iterations $N_{iter} = 10^5$. We have also plotted (represented by continuous line) the theoretical $F$ distribution with the corresponding degrees of freedom i.e. $F(k-1,M-k) = F(4,120)$ for comparison.

We can make the following two observations from this figure:
\begin{itemize}
\item The sampling distribution of $F$ follows the theoretical $F$-distribution curve irrespective of the type of sample distribution. Thus, sample normality is not a prerequisite for  $F$ to be distributed according to $F$-distribution.  

\item Despite a small sample size ($n=25$), the sampling distribution of $F$ approximates well the theoretical curve. Hence, as argued, in practice ANOVA (and other parametric tests) can be applied to approximate the theoretical distribution. Obviously, the approximations will improve with increasing sample size.  

\end{itemize}

We can carry out similar analysis regarding the sampling distribution of the test statistic on real data. However, in practice we typically have only one sample since the subjective or objective experiment is not repeated for obvious reasons. Therefore, to generate the sampling distributions in such scenario, we employ the idea of resampling. Specifically, given two or more samples which are to be compared, we can create randomized versions of these under the assumption that the given samples are similar (i.e. assuming the {\it{null}} hypothesis to be true).  To demonstrate this, we use raw opinion scores from the dataset described in \cite{Pitrey} where a comparison of upscalers was performed at varying compression rates. Since we want to study the sampling distribution of $F$ in ANOVA, we first selected three groups from the said data. These groups represent quality scores of three conditions evaluated by 26 observers. Thus, the group size was 26 ($n_i = n = 26$). Other descriptive properties of the selected groups are summarized in Table \ref{table2} from which we note that none of the groups are normally distributed as indicated by very high or very low kurtosis values and their shapes. In addition, the group variances are similar.  

\begin{table} 
\caption{Description of groups taken from \cite{Pitrey}.}
\label{table2}

\begin{tabular}{|c||c| |c| |c| }
\hline
{}  & group 1 & group 2 & group 3\\
\hline
Mean (MOS) & 5.5769& 7.3846 & 7.3077  \\
\hline
Variance & 3.1338 & 3.2862 & 2.4615\\
\hline
Kurtosis & 1.7971 & 6.9602   & 6.4978 \\
\hline
Shape &\shortstack{unimodal, \\non-symmetric} & \shortstack{bi-modal, \\non-symmetric} & \shortstack{unimodal, \\non-symmetric} \\
\hline
\end{tabular}
\end{table}

\begin{figure*}[!]
\centering
\subfloat[]{\label{three_groups_ANOVA} \includegraphics[width=.4\textwidth]{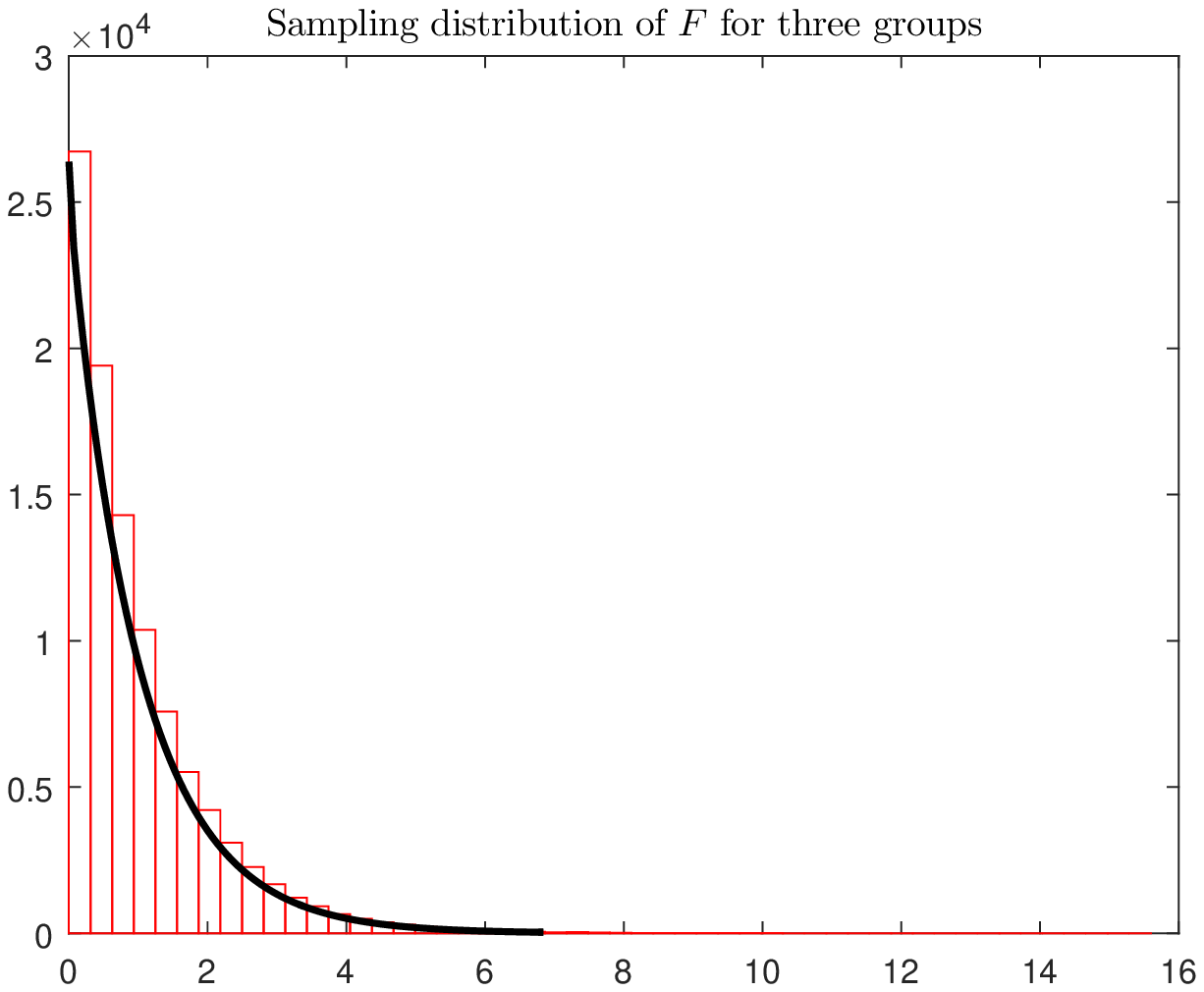}}
\hfil
\subfloat[]{\label{two_groups_t} \includegraphics[width=.4\textwidth]{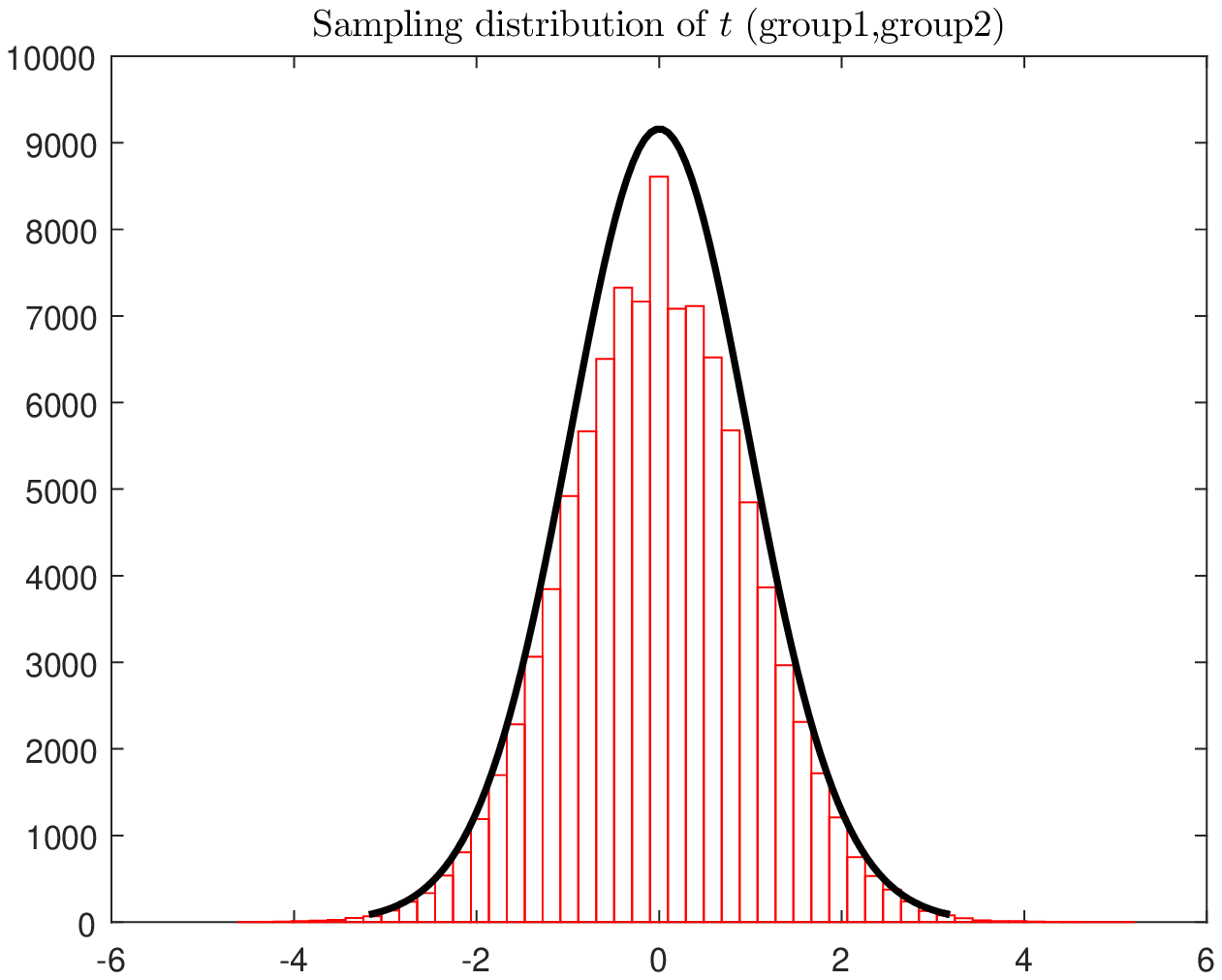}}
\caption{Sampling distribution of $F$ and $t_{pooled}$ values for the groups of data taken from \cite{Pitrey}. The groups are summarized in Table \ref{table2}. In each plot, the continuous curve indicates the corresponding theoretical distribution. Figure best viewed in color.}
\label{sampling distribution of F and t}
\end{figure*}

First, we applied ANOVA to compare the resampled versions of the three groups (we employed $10^5$ randomizations under the {\it{null}} hypothesis) and, the resulting sampling distribution of $F$ values is shown in Figure \ref{three_groups_ANOVA}. As expected, it approximates well the theoretical $F$ distribution. To give another example, we show the sampling distribution of $t_{pooled}$ when comparing group 1 and group 2 using the pooled $t$-test in Figure \ref{two_groups_t}. In this case also, the experimental distribution reasonably follows the theoretical $t$-distribution.

\section{Practical recommendations} \label{recommendations}

Based on the theoretical and experimental analysis in previous sections, it is clear that the application of parametric tests should focus on the consequences of the assumptions in these tests. The practical recommendations towards using the tests are highlighted in the right flow diagram in Figure \ref{dectree}, and are summarized in the following.

Applying normality checks on given data is neither required nor recommended as the CLT provides information about the shape and parameters of the sampling distribution of mean. Instead the more important consideration is whether mean (MOS) adequately represents the desired information from the sample(s). For instance, mean is a useful measure of central tendency in case of many symmetric distributions (not necessarily normal). Moreover, mean is still a practically useful statistic even if there are few outliers (skewness) in the data. In all such cases, parametric tests are practically meaningful for statistical analysis.  

Homogeneity of variance should be exploited to obtain further insights into the data, and therefore not be viewed as a bottleneck for the purpose of statistical testing. To that end, the empirical rule (refer to section \ref{heterogeneous variances}) should be applied to detect the presence of groups/samples that may have very different variances as compared to the remaining ones. If such groups exist, then the corresponding conditions should be revisited to find possible reasons for unequal variance. Consequently, if unequal variance condition is practically reasonable (or such groups can be removed), $t_{pooled}$ or ANOVA can be used. A balanced experimental design (equal sample size) would therefore be preferable in such cases (recall from section \ref{balanced design} both the tests are not affected by unequal variance if group/sample sizes are same). 

The use of nonparametric tests is recommended if mean is not a suitable summary statistic of the data to be analyzed. Note that nonparametric tests should not be used merely because the given data is {\it {nonnormal}}. Rather they should be used to generate the sampling distribution of the desired test statistic. 

In summary, analysis of data pertaining to multimedia quality using mean (average) as a test statistic should focus on experimental design (this includes the selection of challenging content recruiting adequate number of human subjects with possible emphasis on balanced design, conditions to be evaluated, and the final goal of analysis) rather than emphasizing distributional assumptions, equal variance condition or resorting to multiple hypothesis tests. However, if mean is not a suitable test statistic, then nonparametric tests can be used by leveraging the power of computers to construct empirical sampling distribution of the desired test statistic. 

\section{Concluding remarks} \label{concluding remarks}			

Parametric tests provide a theoretical framework for drawing statistical inferences from the data and thus help in formulating well grounded recommendations. However, the application of these tests and interpretation of the results require some care in the light of the assumptions required in these tests. To that end, we revisited the theoretical formulations and clarified the role of the assumption of normality and homogeneity of variance. By analyzing the sampling distribution of the test statistics, we argued that the more appropriate question to be asked before deploying parametric tests is whether the test statistic follows the corresponding distribution or not (instead of the data following any specific distribution). We also emphasized that the said assumptions should not be viewed as constraints on the data. Instead it is more important to focus on their practical implications.   

The presented analysis is particularly relevant in the context of multimedia quality assessment because the said issues have not been emphasized enough in the corresponding literature. We also made practical recommendations in order to avoid the theoretical issues related to multiple hypothesis testing. Even though the targeted application was multimedia quality estimation, the theoretical arguments and the recommendations are expected to be useful in several other areas (such as medical data analysis, information retrieval, natural language processing etc.) where parametric tests are widely used. In order to provide a tool for practical use, a software implementing the said recommendations is also made publicly available\footnote{https://sites.google.com/site/narwariam/home/research}.





\bibliographystyle{IEEEtran}
\bibliography{refs_apa}
%
%
%

\end{document}